\documentclass[twocolumn,showpacs,preprintnumbers,amsmath,amssymb]{revtex4}
\usepackage{amssymb}

\usepackage{graphicx}
\usepackage{dcolumn}
\usepackage{bm}

\begin{document}

\title{Experimental demonstration of photonic entanglement collapse and revival}
\author{Jin-Shi Xu}
 \affiliation{Key Laboratory of Quantum Information,
  University of Science and Technology
  of China, CAS, Hefei, 230026, People's Republic of China}
 \author{Chuan-Feng Li$\footnote{email: cfli@ustc.edu.cn}$}
\affiliation{Key Laboratory of Quantum Information,
  University of Science and Technology
  of China, CAS, Hefei, 230026, People's Republic of China}
\author{Ming Gong}
\affiliation{Key Laboratory of Quantum Information,
University of Science and Technology
of China, CAS, Hefei, 230026, People's Republic of China}
\author{Xu-Bo Zou$\footnote{email: xbz@ustc.edu.cn}$}
 \affiliation{Key Laboratory of Quantum Information,
  University of Science and Technology
  of China, CAS, Hefei, 230026, People's Republic of China}
\author{Cheng-Hao Shi}
 \affiliation{Key Laboratory of Quantum Information,
  University of Science and Technology
  of China, CAS, Hefei, 230026, People's Republic of China}
\author{Geng Chen}
\affiliation{Key Laboratory of Quantum Information,
  University of Science and Technology
 of China, CAS,  Hefei, 230026, People's Republic of China}
 \author{Guang-Can Guo}
\affiliation{Key Laboratory
of Quantum Information, University of Science and Technology of
China, CAS, Hefei, 230026, People's Republic of China}
\date{\today }
\begin{abstract}
We demonstrate the collapse and revival features of the entanglement
dynamics of different polarization-entangled photon states in a
non-Markovian environment. Using an all-optical experimental setup,
we show that entanglement can be revived even after it suffers from
sudden death. A maximally revived state is shown to violate a Bell's
inequality with 4.1 standard deviations which verifies its quantum
nature. The revival phenomenon observed in this experiment provides
an intriguing perspective on entanglement dynamics.
\end{abstract}

\pacs{03.67.Mn, 03.65.Ud, 03.65.Yz}
\maketitle

Quantum entanglement, which is a kind of counterintuitive nonlocal
correlation, is fundamental in quantum physics both for its
essential role in understanding the nonlocality of quantum mechanics
\cite{Einstein35,Bell64} and its practical application in quantum
information processing \cite{Nielsen00,Bennett00}. However,
entanglement will become degraded due to the unavoidable interaction
with the environment \cite{Zurek03,Schlosshauer05}. Recently, the
dynamics of entanglement in different noise channels has attracted
extensive interests
\cite{Zyczkowski01,Yu04,Roszak06,Yu06,Yonac06,Yonac08,Bellomo07,Bellomo08,Lopez08,Almeida07,Laurat07,Yu09}.
Surprisingly, the evolution of entanglement may possess some
distinct properties. It has been shown that entanglement between two
particles evolved in independent reservoirs may disappear completely
at a finite time in spite of the asymptotical coherence decay of
single particle \cite{Zyczkowski01,Yu04,Roszak06,Yu06}. This
phenomenon, termed as entanglement sudden death (ESD) \cite{Yu06},
has been experimentally observed in Markovian environments
\cite{Almeida07,Laurat07} (for a review see \cite{Yu09} and
references therein). Moreover, different from the irreversible
disentanglement process in the Markovian environment, non-Markovian
noise with memory effect may contribute to the revival of
entanglement even after ESD occurs
\cite{Zyczkowski01,Yonac06,Yonac08,Bellomo07,Bellomo08}. Here, we
experimentally investigate the collapse and revival of entanglement
of two photons with one of them passing through a birefringent
non-Markovian environment, which is simulated by a special designed
Fabry-Perot (FP) cavity followed by quartz plates. We observe the
revival of entanglement after it suffers from sudden death. A
maximally revived state is shown to violate a Bell's inequality with
4.1 standard deviations which disproves its local realistic
description.

Consider one of the maximally entangled polarization states
$|\phi\rangle=1/\sqrt{2}(|HH\rangle_{a,b}+|VV\rangle_{a,b})$, where
$H$ and $V$ represent the horizontal and vertical polarizations,
respectively. The subscripts $a$ and $b$ denote the different paths
of the photons. If the photon in mode $b$ passes through
birefringent crystals (quartz plates) with the optic axes set to be
horizontal, the final polarization state of the two photons can be
written as the following reduced density operator \cite{Berglund00}
\begin{equation}
\rho=\frac{1}{2}(|HH\rangle\langle HH|+|VV\rangle\langle
VV|+\kappa_{b}^{\ast}|HH\rangle\langle
VV|+\kappa_{b}|VV\rangle\langle HH|),\label{pure}
\end{equation}
where the decoherence parameter $\kappa_{b}=\int f(\omega_{b})
\exp(i\alpha\omega_{b}) \mathrm{d}\omega_{b}$, with $f(\omega_{b})$
denoting the amplitude corresponding to the frequency $\omega_{b}$
of the photon in mode $b$ and being normalized as $\int
f(\omega_{b})\mathrm{d}\omega_{b}=1$. In our case, $\alpha=L\Delta
n/c$ where $L$ is the thickness of quartz plates and $c$ represents
the velocity of the photon in the vacuum. $\Delta n=n_{o}-n_{e}$ is
the difference between the indices of refraction of ordinary
($n_{o}$) and extraordinary ($n_{e}$) light. Generally, the
frequency spectrum $f(\omega_{b})$ is peaked at some central value
$\omega_{0}$ with a finite width $\sigma$. For example, the Gaussian
function like frequency distribution of the photon can be written as
$f(\omega_{b})=(2/\sqrt{\pi}\sigma)\exp(-4(\omega_{b}-\omega_{0})^{2}/\sigma^{2})$.
The photon with different frequency will experience different
relative phase between horizontal and vertical polarization states
according to the relationship of $\alpha\omega_{b}$.  Therefore, the
value of the off-diagonal element of the final density matrix
$\kappa_{b}=\exp(-\alpha^{2}\sigma^{2}/16+i\alpha\omega_{0})$
degrades exponentially and the final state turns to the maximally
mixed state without any entanglement remained after long enough
interaction time (i.e., with $L$ long enough). However, if the
Gaussian frequency distribution is filtered by a FP cavity with
carefully selected parameters, the spectrum will exhibit discrete
distribution and it can be written as
$f(\omega_{b})=\displaystyle\sum_{j=1}^N
A_{j}\frac{2}{\sqrt{\pi}\sigma_{j}}\exp(-4(\omega_{b}-\omega_{j})^{2}/\sigma_{j}^{2})$,
where $A_{j}$ are the relative amplitudes of these finite $N$
Gaussian functions with the central frequencies $\omega_{j}$ and
frequency widths $\sigma_{j}$. As a result, the off-diagonal element
becomes
$\kappa_{b}=\displaystyle\sum_{j=1}^{N}A_{j}\exp(-\alpha^{2}\sigma_{j}^{2}/16+i\alpha\omega_{j})$.
During the evolution, the overall relative phase may refocus and
off-diagonal elements reappear.

For two-qubit states, entanglement can be quantified by the
concurrence \cite{Wootters98}, which is given by
\begin{equation}
C=\textrm{max}\{0,\Gamma\}, \label{concurrence}
\end{equation}
where
\begin{equation}
\Gamma=\sqrt{\chi_{1}}-\sqrt{\chi_{2}}-\sqrt{\chi_{3}}-\sqrt{\chi_{4}},\label{gamma}
\end{equation}
and the quantities $\chi_{j}$ are the eigenvalues in decreasing
order of the matrix
$\rho(\sigma_{y}\otimes\sigma_{y})\rho^{\ast}(\sigma_{y}\otimes\sigma_{y})$
with $\sigma_{y}$ denoting the second Pauli matrix and $\rho^{\ast}$
corresponding to the complex conjugate of $\rho$  in the canonical
basis $\{|HH\rangle,|HV\rangle,|VH\rangle,|VV\rangle\}$. According
to equations (1) and (2), we can get the concurrence as
$C=|\kappa_{b}|$. Therefore, the revival of $|\kappa_{b}|$ will lead
to the revival of entanglement
\cite{Zyczkowski01,Yonac06,Yonac08,Bellomo07,Bellomo08}.

Actually, the dynamics of entanglement in bipartite quantum systems
is sensitive to initial conditions \cite{Roszak06,Yu06}.  ESD occurs
for some special states of two particles coupled to independent
amplitude decay channels whereas the entanglement will
asymptotically disappear in phase-damping channels \cite{Almeida07}.
However, it has been shown that under strong partial pure dephasing,
ESD can also occur for certain states \cite{Roszak06}. Now, we
consider the entanglement dynamics of a partially entangled input
state. This initial state is prepared by implementing $\sigma_{x}$
operation on the photon in mode $a$ of the maximally entangled state
$|\phi\rangle$ and then further dephasing it in $H/V$ bases. The
photon in mode $b$ then passes through the same non-Markovian
environment mentioned above. The final state becomes
\begin{equation}
\rho=\frac{1}{4}\left(
\begin{array}{cccc}
1 & \kappa_{b}^{\ast} & \kappa_{a}^{\ast} & -\kappa_{a}^{\ast}\kappa_{b}^{\ast} \\
\kappa_{b} & 1& \kappa_{a}^{\ast}\kappa_{b} & -\kappa_{a}^{\ast} \\
\kappa_{a}& \kappa_{a}\kappa_{b}^{\ast}& 1&-\kappa_{b}^{\ast} \\
-\kappa_{a}\kappa_{b} & -\kappa_{a}& -\kappa_{b}&1%
\end{array}%
\right),\label{mix}
\end{equation}
where $\kappa_{a}$ is the decoherence parameter in mode $a$. For the
simplified case where $\kappa_{a}$ and $\kappa_{b}$ are set to be
real in equation (4), the concurrence is therefore given by
$C=\textrm{max}\{0,\frac{1}{2}(\kappa_{a}+\kappa_{a}\kappa_{b}+\kappa_{b}-1)\}$.
We can see that ESD \cite{Yu06} occurs when
$\kappa_{b}=\frac{1-\kappa_{a}}{1+\kappa_{a}}$. However, in such a
non-Markovian environment, $\kappa_{b}$ will be revived with the
increasing of interaction time and entanglement revival from sudden
death can be realized.

\begin{figure}[tbph]
\begin{center}
\includegraphics [width= 2.7in]{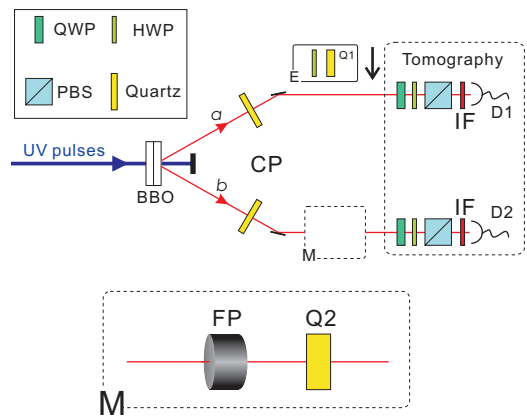}
\end{center}
\caption{(Color online). Experimental setup. Degenerate
polarization-entangled photons are generated by the process of
spontaneous parametric down-conversion in the two-crystal geometry
type I BBO crystals. These two photons pass through quartz plates
(C) to compensate the birefringence in BBO. The half-wave plate
(HWP) and quartz plates (Q1) in the solid pane E are inserted into
mode $a$ to prepare the partially entangled input state. The dashed
pane M which contains a Fabry-Perot (FP) cavity and quartz plates
(Q2) in mode $b$ is used to simulate a non-Markovian decoherence
environment (the absorption of quartz plates is negligible and there
is not significant change in the total coincidence rate by
increasing Q2). After passing through quarter-wave plates (QWP),
half-wave plates and polarization beam splitters (PBS) which allow
tomographic reconstruction of the density matrix, both photons are
then registered by single-photon detectors (D1 and D2) equipped with
3 nm interference filters (IF).} \label{fig:setup}
\end{figure}

Our experimental setup is shown in Figure 1. Ultraviolet (UV) pulses
are frequency doubled from a mode-locked Ti:sapphire laser centered
at 780 nm with 130 fs pulse width and 76 MHz repetition rate. These
UV pulses prepared to be $45^{\circ}$  linearly polarized are then
focused into two identically cut type-I beta-barium-borate (BBO)
crystals with their optic axes aligned in mutually perpendicular
planes to produce degenerate polarization-entangled photon pairs
\cite{Kwiat99}. Inside the crystals, an UV photon may spontaneously
convert into a photon pair with vertical polarizations in one
crystal or horizontal polarizations in the other. By carefully
compensating the birefringence between $H$-polarization and
$V$-polarization components in the two-crystal geometry BBO with
quartz plates (C), we can get the maximally polarization-entangled
state $|\phi\rangle=1/\sqrt{2}(|HH\rangle+|VV\rangle)$ with high
visibility \cite{Xu06}.

The half-wave plate with the optic axis set at $22.5^{\circ}$ and
quartz plates (Q1) set to be horizontal in the solid pane E in Fig.
1 are inserted into mode $a$ in the case of considering the
entanglement evolution of the partially entangled state. A FP cavity
followed by quartz plates (Q2) which locates in the dashed pane M is
used to simulate the non-Markovian environment. This FP cavity is a
0.2 mm thick quartz glass with coating films (reflectivity 90\% at
wavelengths around 780 nm) on both sides. Although wavelengths
within the reflective band of the FP cavity are reflected,
wavelengths for which the cavity optical thickness is equal to an
integer multiple of half wavelengths are completely transmitted due
to the effect of multi-beam interference. We then use the standard
quantum state tomography with the usual 16 coincidence measurements
to character the density matrices of final states \cite{James01}. By
employing the maximum likelihood estimation \cite{James01}, we get
non-negative definite density matrices and then calculate the
concurrences. The 3 nm (full width at half maximum) interference
filters (IF) are used not only to increase the coherence time of the
photons, but also to reduce the number of spectral lines transmitted
through the FP cavity.
\begin{figure}[tbph]
\begin{center}
\includegraphics [width= 3in]{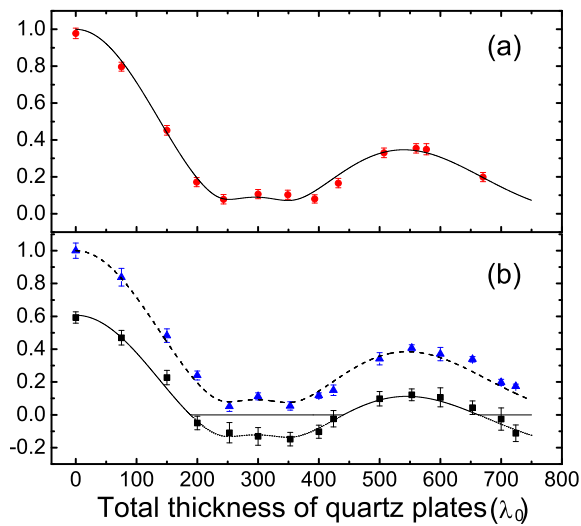}
\end{center}
\caption{(Color online). Experimental results for the non-Markovian
evolution. (a) Entanglement dynamics of the maximally entangled
input state. Red dots are the experimental results of $\Gamma$. The
solid line and dotted line are the theoretical prediction of the
concurrence and $\Gamma$ given by equation (2) and (3),
respectively, which are completely overlapped in this case and only
the solid line can be seen. (b) Entanglement evolution of the
partially entangled input state and the evolution of the degree of
polarization $P$ of the photon in mode $b$. Black squares represent
the experimental values of $\Gamma$. The theoretical fitting of
concurrence (solid line) is set to 0 when the quantity $\Gamma$
(dotted line) becomes less than 0. Blue regular triangles represent
the experimental results of $P$. The dashed line is the theoretical
fitting given by $|\kappa_{b}|$. $\lambda_{0}$=0.78 $\mu$m.}
\label{fig:concurrence}
\end{figure}

Fig. 2(a) shows the evolution of the concurrence and the quantity
$\Gamma$ of the maximally entangled input state, as a function of
the thickness of Q2 ($L$),  while Fig. 2(b) displays the evolution
of the partially entangled input state and the degree of
polarization $P$ of the photon in mode $b$. The concurrence of the
maximally entangled state we prepared is about 0.978. The evolution
of concurrence of such input state is the same as that of the
quantity $\Gamma$. We can see from Fig. 2(a), when $L$ is increased
to $243\lambda_{0}$ ($\lambda_{0}$=0.78 $\mu$m), the concurrence
drops nearly to zero. After keeping on an almost flat section, the
concurrence begins growing with further increasing $L$ and reaches
its maximal value 0.354 at about $560\lambda_{0}$. When we consider
the entanglement dynamics of the partially entangled input state,
the solid pane E in Fig. 1 is inserted in mode $a$ and the thickness
of Q1 is set to be $117\lambda_{0}$. It can be seen from Fig. 2(b),
when Q2 reaches about $189\lambda_{0}$, $\Gamma$ (black squares)
become less than zero and the concurrence is given by zero according
to equation (2), which clearly shows the phenomenon of ESD
\cite{Yu06}. After a completely dark period, due to the refocusing
of the overall relative phase, $\Gamma$ become positive again and
the revival of entanglement from sudden death is realized when
$L=440\lambda_{0}$. With further increasing Q2, the concurrence
reaches its maximal value about 0.11 at around $540\lambda_{0}$. We
can see that ESD occurs again when Q2 is increased to
$663\lambda_{0}$. The solid lines are the corresponding fittings of
the concurrence, given by equation (2), while the dotted lines are
the theoretical fittings of $\Gamma$ using equation (3). They
completely overlap in Fig. 2(a) and only the solid line can be seen.

In order to demonstrate the difference between the dynamics of
entanglement and single-photon coherence, we also show in Fig. 2(b)
the evolution of the degree of polarization $P$ of the photon in
mode $b$. It is defined as $P=\sqrt{\langle s_{1}\rangle^{2}+\langle
s_{2}\rangle^{2}+\langle s_{3}\rangle^2}$, where these three Stokes
parameters are calculated as $\langle s_{1}\rangle=2\langle
H|\rho_{b}|H\rangle-1$, $\langle s_{2}\rangle=\langle
H|\rho_{b}|V\rangle+\langle V|\rho_{b}|H\rangle$ and $\langle
s_{3}\rangle=i(\langle H|\rho_{b}|V\rangle-\langle
V|\rho_{b}|H\rangle)$ \cite{Berglund00,Mandel95,Alodjants99} and
$\rho_{b}$ is the density matrix of photon in mode $b$ triggered by
the photon in mode $a$ in the horizontal polarization. Therefore,
$P=|\kappa_{b}|$. Blue regular triangles represent the experimental
results and the dashed line is the corresponding theoretical
fitting. The residual value of $P$ is about 0.24 when ESD occurs,
which shows that the particular property of ESD occurs faster than
the single-photon decoherence.

In our experimental fittings, the frequency distribution in mode $a$
defined by the 3 nm interference filter is treated as the Gaussian
wave function with the central wavelength 780 nm and $\kappa_{a}$ is
calculated to be about 0.607. While the discrete frequency
distributions in mode $b$ is treated as three Gauss like wave
packets ($\omega_{j}$) centered at 778.853, 780.160 and 781.459 nm
with relative probabilities ($A_{j}$) of 0.37, 0.44 and 0.19,
respectively \cite{Note}. These spectrum widths ($\sigma_{j}$) in
mode $b$ are identically fitted to 0.9 nm for the case of Fig. 2(a)
and 0.85 nm for the case of Fig. 2(b). This slight difference is due
to the different reflectivity of the used FP cavity in these two
cases. In our experiment, $\Delta n$ is treated as the constant of
0.01 for the small frequency distribution. We find good agreement
between the experimental results and theoretical fittings. Errors of
the experimental results come mainly from the random fluctuation of
each measured coincidence counts and the uncertainties in aligning
the wave plates \cite{James01}.

\begin{figure}[tbph]
\begin{center}
\includegraphics [width= 2.0in]{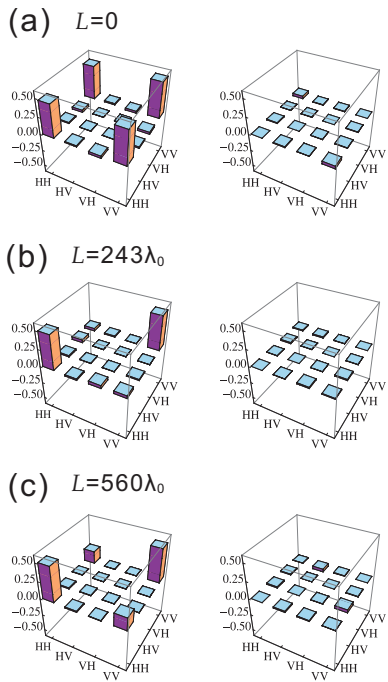}
\end{center}
\caption{(Color online). Graphical representations of density
matrices for three states during the evolution of the maximally
entangled input state. The first and second columns represent the
real and imaginary parts of the density matrices, respectively. (a)
The density matrix of the initial entangled state with $L=0$.  (b)
The maximal decoherence state with $L=243\lambda_{0}$.  (c) The
maximally revived state with $L=560\lambda_{0}$.}
\label{fig:density}
\end{figure}

The entanglement revival phenomenon can also be seen from the
reappearance of the off-diagonal elements in the density matrix of
the final state. Fig. 3 shows the case of the evolution of the
maximally entangled input state. From the real parts (first column
in Fig. 3), we can see that the initial entangled state has the
largest off-diagonal elements which become almost vanished at
$L=243\lambda_{0}$, i.e., the state has nearly lost its coherence
completely. The maximal reappearance of the off-diagonal elements is
achieved when $L=560\lambda_{0}$. This result is consistent with the
revival of concurrence, as shown in Fig. 2(a).

Nonlocality, which is the essential characteristic of quantum
mechanics, has stimulated great interests. As we show below, our
maximal revival entangled state can violate a suitable Bell's
inequality which disproves its local realistic description. In
particular, according to the Clauser-Horne-Shimony-Holt (CHSH)
inequality \cite{Clauser69}, $S\leq2$  for any local realistic
theory, where
\begin{equation}
S=E(\theta_{1},\theta_{2})+E(\theta_{1},\theta_{2}^{'})+E(\theta_{1}^{'},\theta_{2})
-E(\theta_{1}^{'},\theta_{2}^{'}), \label{CHSH}
\end{equation}
with $E(\theta_{1},\theta_{2})$ representing the coefficient for
joint measurement. $\theta_{1}$ (or $\theta_{1}^{'}$) is the linear
polarization setting for the photon in mode $a$ and $\theta_{2}$ (or
$\theta_{2}^{'}$)  is the setting for the photon in mode $b$. We set
$\theta_{1}=-86.25^{\circ}$, $\theta_{1}^{'}=60.75^{\circ}$,
$\theta_{2}=-85.5^{\circ}$ and $\theta_{2}^{'}=76.5^{\circ}$, which
are calculated from the maximally revived density matrix to maximize
the quantum mechanics prediction of $S$. We get $S=2.045\pm0.011$
which violates the local realism limit 2 by about 4.1 standard
deviations and clearly shows the quantum nature of the revived
state. This result is deduced from the raw data without any
corrections and the uncertainty is due to counting statistics.

In summary, our work shows the features of entanglement collapse and
revival of different input states in a non-Markovian environment. To
our best knowledge, it is the first time to use a FP cavity for the
introduction of non-Markovian features into the noise channel. The
non-Markovian environment acting via the FP cavity and quartz plates
on only one of the photons retains a memory of the two-photon state
at a given time and then later passes this information back into
this state (relative phase refocusing), which leads to collapse and
revive the entanglement in the two-photon system. Future work would
concern on the issue of non-local recovery with a non-Markovian
environment acting on both photons. Moreover, a maximally restored
state can violate the CHSH inequality with 4.1 standard deviations
which confirms its quantum nature. The revival phenomenon observed
in this experiment provides an intriguing perspective on
entanglement dynamics.

We thank J. H. Eberly and F.-W. Sun for helpful discussion. This
work was supported by National Fundamental Research Program (Grant
No. 2006CB921900), also by National Natural Science Foundation of
China (Grant No. 10674128, 10874162 and 60621064).

\end{document}